\documentclass[
reprint,
twocolumn,
aps,
prr,
citeautoscript,
showpacs,
showkeys,
groupedaddress,
superscriptaddress,
twocolumn
amsmath,
amssymb,
]{revtex4-2}

\usepackage{graphicx}
\usepackage{dcolumn}
\usepackage{bm}
\usepackage{ulem}

\usepackage[hidelinks]{hyperref}
\hypersetup{
	colorlinks,
	linkcolor={blue!50!black},
	citecolor={blue!50!black},
	urlcolor={blue!80!black}
}

\usepackage{siunitx}
\usepackage{xcolor}
\usepackage{braket}
\usepackage{enumitem}

\usepackage{mathtools}
\newcommand{\Li}{\textsuperscript{6}Li }

\newcommand{\rom}[1]{\uppercase\expandafter{\romannumeral #1\relax}}

\usepackage{hyperref}

\begin{document}
	
\title{Engineering a Bound State in the Continuum via Quantum Interference}

\author{Alexander Guthmann}
\affiliation{Department of Physics and Research Center OPTIMAS, RPTU University Kaiserslautern-Landau, 67663 Kaiserslautern, Germany}

\author{Louisa Marie Kienesberger}
\affiliation{Department of Physics and Research Center OPTIMAS, RPTU University Kaiserslautern-Landau, 67663 Kaiserslautern, Germany}

\author{Felix Lang}
\affiliation{Department of Physics and Research Center OPTIMAS, RPTU University Kaiserslautern-Landau, 67663 Kaiserslautern, Germany}

\author{Eleonora Lippi}
\affiliation{Department of Physics and Research Center OPTIMAS, RPTU University Kaiserslautern-Landau, 67663 Kaiserslautern, Germany}

\author{Artur Widera}
\email[]{widera@rptu.de}
\affiliation{Department of Physics and Research Center OPTIMAS, RPTU University Kaiserslautern-Landau, 67663 Kaiserslautern, Germany} 

\date{\today}

\begin{abstract}
Quantum mechanical interaction potentials typically support either localized bound states below the dissociation threshold or delocalized scattering states above it. 
While bound states are energetically isolated, scattering states embed a quantum system in a continuum of environmental modes, making dissipation and loss intrisic features of open quantum systems. 
A striking exception are bound states in the continuum (BICs), which remain localized despite lying within the scattering continuum due to destructive interference.
It was predicted that such states can arise from the interference of two Feshbach resonances coupled to a common continuum, 
yet this mechanism has remained experimentally inaccessible in genuine quantum systems. 
Here we demonstrate the formation of such an interference-stabilized state in ultracold collisions of \Li atoms by coherently coupling two tunable Feshbach resonances using Floquet engineering. At a critical parameter point, both elastic and inelastic coupling to the continuum vanish, yielding a molecular state above the dissociation threshold. 
Loss spectroscopy, quench dynamics, and rf-photoassociation directly reveal the resulting decoupling from scattering states. Our observations are quantitatively captured by full coupled-channel calculations and a minimal non-Hermitian model, identifying a Friedrich–Wintgen BIC. 
Our results establish quantum interference as a powerful mechanism for controlling openness in quantum matter and for engineering non-Hermitian Hamiltonians.
\end{abstract}

\maketitle

The interaction of a quantum system with its environment is both unavoidable and
fundamental, governing dissipation, decoherence, and irreversibility \cite{Breuer2007}. While coupling to
external continua is often treated as a limitation, it can also give rise to counterintuitive interference effects that suppress decay rather than enhance it.
\begin{figure}
	\centering
	\includegraphics[scale=1.0]{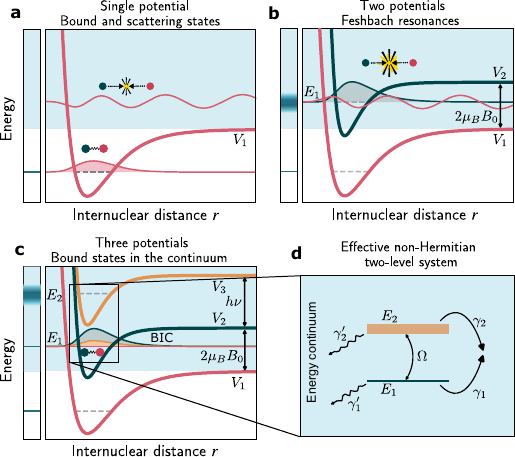}
	\caption{\textbf{Emergence of a Friedrich–Wintgen BIC.} \textbf{(a)}  Energy spectrum and wavefunctions for a molecular potential $V_1$. Below the molecular dissociation threshold $V_1(r\to\infty)$, discrete bound states exist with wavefunctions localized in the potential well. Above the threshold, the spectrum becomes continuous and the solutions describe delocalized scattering states. \textbf{(b)} For alkali atoms, spin-dependent coupling between the open channel $V_1$ (triplet) and a closed channel $V_2$ (singlet) gives rise to Feshbach resonances, producing finite-lifetime resonant states at energy $E_1$ that mixes closed localized (blue) and open scattering (red) channels. Since the singlet and triplet channels possess different magnetic moments, an external magnetic field $B_0$ shifts their relative energies and thus allows the resonance position to be tuned. \textbf{(c)} In the presence of a second closed channel $V_3$ (yellow), destructive interference between two individual Feshbach resonances at energies $E_{1,2}$ can decouple one of them from the continuum if the detuning $\delta=E_1-E_2$ fulfills the BIC condition. In this case, the delocalized open channel contribution (red) vanishes completely and  only spatially localized closed channel contributions (blue and yellow) remain, forming a Friedrich–Wintgen BIC above molecular dissociation threshold. Floquet engineering provides the modulation frequency $\nu$ as a continuous tuning parameter for $\delta$. \textbf{(d)} Effective non-Hermitian two level system modeling the Friedrich–Wintgen BIC mechanism. Two states with energies $E_{1,2}$ are coupled to each other with Rabi-frequency $\Omega$ and to a common continuum with coherent rates $\gamma_{1,2}$; rates $\gamma'_{1,2}$ take into account incoherent decay. Direct and continuum-mediated coupling pathways can interfere destructively canceling the effective decay of one resonance and yielding a non-decaying bound state embedded in the continuum.}
		\label{fig:fig1}
\end{figure}
One striking example is a bound state embedded in the continuum.
Usually in quantum mechanics, a particle's energy and the shape of the interaction potential $V(r)$ determine whether the particle forms a bound state or undergoes scattering.
For a potential well with a finite threshold, bound states typically exist only for energies below this threshold, while higher-energy solutions correspond to delocalized, unbound scattering states [Fig.~\ref{fig:fig1} (a)].
Such scattering states can give rise to scattering resonances, including Feshbach resonances [Fig.~\ref{fig:fig1}(b)] that play a central role in nuclear and ultracold atom physics \cite{Chin2010}.
Bound states in the continuum (BICs) defy this conventional distinction between bound and scattering states. 
They are solutions to the Schrödinger equation that remain spatially localized and normalizable, even though their energy lies within the continuum of scattering states above the dissociation threshold. 
This is in stark contrast to surrounding non-square integrable continuum solutions. 
Established methods to achieve decoupling include reducing the density of states at the relevant energy, i.e. forming energy gaps \cite{Breuer2007} as well as engineering symmetry-protected, e.g. through selection rules \cite{Cohen-Tannoudji08}, or topology-protected states \cite{Hasan10}, which are all actively pursued in current research for quantum computing using noisy intermediate-scale quantum devices \cite{Preskill18}. Our scattering control adds a novel way to decouple a bound state via quantum interference in open quantum systems.

Originally predicted in the early days of quantum mechanics by von Neumann and Wigner through an explicitly constructed potential~\cite{vNeuman1929, Stillinger1975}, BICs have since been recognized as a general interference phenomenon~\cite{Hsu2016, Sadreev2021}. 
In 1985 Friedrich and Wintgen proposed that a BIC in quantum systems could arise from the interference between two Feshbach resonances \cite{Friedrich1985}, as illustrated in Fig.~\ref{fig:fig1}(c). 
They suggested that by varying external tuning parameters one can induce an avoided crossing between Feshbach resonances, giving rise to a BIC.
In the context of two colliding atoms, such a BIC corresponds to a counterintuitive molecular bound state above the dissociation threshold.

\begin{figure*}
	\centering
	\includegraphics[scale=0.95]{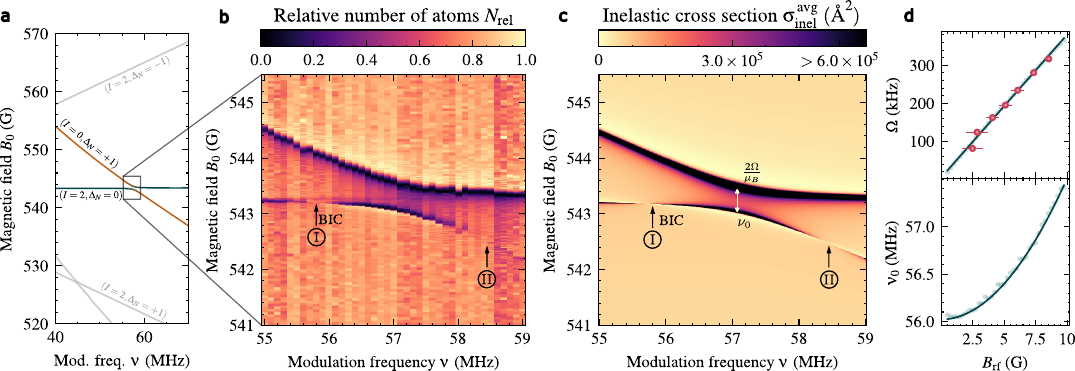}
	\caption{\textbf{Appearance of a BIC at an avoided crossing of Feshbach resonances.} \textbf{(a)} Modulation-frequency dependence of Floquet–Feshbach resonances. Floquet modulation dresses the molecular states and induces additional Feshbach resonances that tune differently with modulation frequency depending on the underlying molecular state causing the resonance. When two such resonances come close an avoided crossing forms. \textbf{(b)} Experimental atom-loss spectroscopy across the avoided crossing between the $(I=2,\Delta_N=0)$ and $(I=0,\Delta_N=+1)$ Floquet–Feshbach resonances. The lower loss branch vanishes in two distinct regions: \rom{1} around a modulation frequency of $\qty{55.8}{MHz}$ and \rom{2} near $\qty{58.5}{MHz}$. The BIC forms in region \rom{1}. The faint horizontal feature between the two branches stems from the static $I=2$ resonance and is an artifact of the experimental sequence. \textbf{(c)} Coupled-channel calculations of the thermally averaged inelastic scattering cross section $\sigma^\text{avg}_\text{inel}$ reproduce the observed avoided crossing and identify the same regions \rom{1} and \rom{2}. \textbf{(d)} Dependence of the coupling strength $\Omega$ (upper panel) and numerically calculated position of the avoided crossing $\nu_0$ (lower panel) on the modulation amplitude $B_\text{rf}$. The coupling strength increases linearly with $B_\text{rf}$ (red dots: experiment; faint blue dots: coupled-channel calculations; blue line: linear fit, slope $\qty{37.65}{kHz/G}$. Strong modulation also induces an AC-Zeeman shift that quadratically shifts the crossing's center frequency $\nu_0$ (faint blue dots: numerical data; blue line: quadratic fit $\nu_0 = \alpha B_\text{rf}^2+\beta$, with: $\alpha=\qty{0.0149}{MHz/G^2}$ and $\beta=\qty{56.04}{MHz}$).}
	\label{fig:fig2}
\end{figure*}

Experimental implementations of BICs have so far been predominantly realized in classical systems such as photonic crystals~\cite{Hsu2013}, photonic waveguides~\cite{Plotnik2011, Crespi2015}, and acoustic systems~\cite{PARKER196662}. 
In these platforms, BICs typically correspond to single-particle wave modes that are localized in momentum space, often protected by symmetry and fixed by the underlying geometry of the structure.
In contrast, a BIC formed in ultracold atomic collisions corresponds to a genuinely two-body molecular bound state coexisting with the continuum of scattering states.
Beyond their fundamental significance, realizations of BICs have driven key technological advances in optics and photonics, enabling ultra-high-$Q$ resonators for precision applications~\cite{Kang2023}, enhanced sensing platforms~\cite{QIU2025}, and the engineering of advanced metasurfaces~\cite{Liu2019}. 

More recently, topological corner states in ultracold atoms using synthetic momentum lattices have been shown to host higher-order topological BICs \cite{dong2025}. 
Recent theoretical advances also include studies of BICs in dissipative driven lattices \cite{Zhao2025} and predictions of BICs in spin-orbit-coupled atomic systems \cite{Kartashov2017}.
Earlier experiments on autoionization and laser-induced continuum structure (LICS) revealed very narrow spectral features arising from interference \cite{Neukammer1985, Bohmer2002}.

Yet, despite these broad developments, a quantum-mechanical realization of the original Friedrich–Wintgen mechanism, based on destructive interference between two continuously tunable Feshbach scattering resonances, has remained experimentally elusive.
Although their work established the theoretical framework, it did not identify the specific control parameters required to achieve tunability and to observe the avoided crossing between two Feshbach resonances. 
Laser-based cold-atom accelerators, which offer control over the collision energy, provide one potential route~\cite{Rakonjac2012}, and a Friedrich–Wintgen BIC has been predicted for collisions of $^{87}$Rb atoms~\cite{Chilcott2024}.
Theoretical work also shows that a BIC can be engineered through coherent optical coupling near a magnetic Feshbach resonance~\cite{Deb2014}.
However, no direct observation has been reported to date. 

Here, we realize a Friedrich–Wintgen BIC in an ultracold atomic gas of \Li through Floquet engineering of Feshbach resonances. 
To this end, we exploit the recently accomplished control mechanism based on strong magnetic-field modulation generating additional Feshbach scattering resonances with tunable positions and widths \cite{Guthmann2025}.
Modulation-induced modification of scattering properties has been investigated in a range of theoretical and experimental works \cite{tscherbul_rf-field-induced_2010, Papoular2010, Beaufils2010, owens_creating_2016, Holthaus2016, Eckardt2017, Ding2017, smith_inducing_2015, Dauer2025, ballu2025}. 
In our approach for the Friedrich–Wintgen BIC, the modulation frequency $\nu$ serves as a tunable control parameter, enabling the realization of a BIC at experimentally accessible collision energies.
This establishes an novel class of states bound due to quantum interference in addition to those bound by energy gaps or topological invariance.

\begin{figure*}
	\centering
	\includegraphics[scale=1.0]{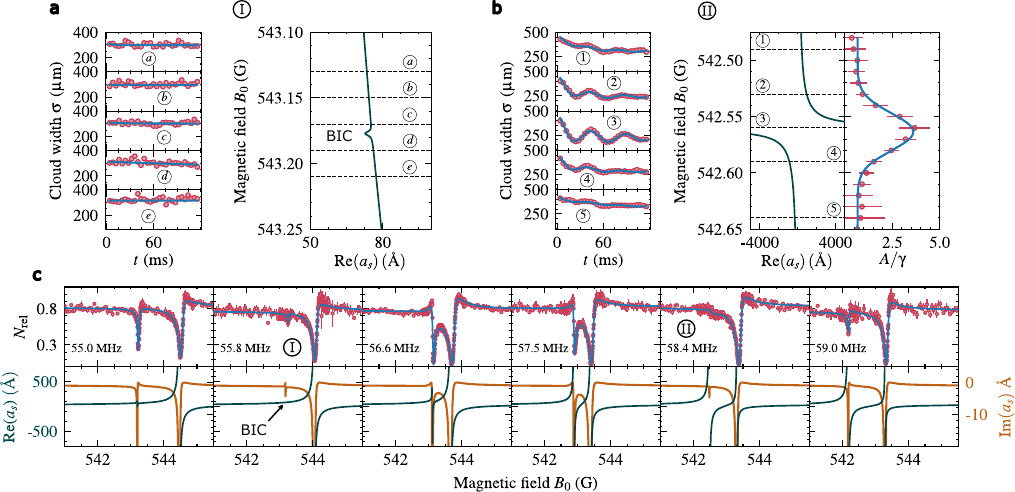}
	\caption{\textbf{Characterization of the BIC.}  \textbf{(a)}, \textbf{(b)} Response of the gas to a sudden trap quench recorded at points \rom{1} \textbf{(a)} and \rom{2} \textbf{(b)}. At point \rom{1}, the pole in $\text{Re}(a_s)$ vanishes, and the cloud exhibits no collective motion, demonstrating that the state is decoupled from the continuum forming a BIC. In contrast at \rom{2}, the resonance pole in $\text{Re}(a_s)$ drives pronounced collective breathing oscillations, which can be fitted with an exponentially decaying sine with amplitude $A$ and decay coefficient $\gamma$ (see Methods). \textbf{(c)} Cuts through the avoided crossing at selected modulation frequencies. The upper panels show the relative remaining atom number after modulation; the lower panels show the real (blue) and imaginary (ochre) parts of the $s$-wave scattering length $a_s$, associated with elastic and inelastic processes, respectively. At point \rom{1} both  the pole in $\text{Re}(a_s)$ and the extrema in $\text{Im}(a_s)$ vanish, indicating decoupling from the continuum and formation of a BIC. At point \rom{2} only the imaginary part is suppressed, yielding resonant elastic scattering with suppressed losses.}
	\label{fig:fig3}
\end{figure*}

\section{Effective Non-Hermitian Model}

The Friedrich–Wintgen BIC mechanism can be understood within a minimal model of two discrete states coupled to a common continuum [Fig.~\ref{fig:fig1}(d)].
This situation is effectively described by a non-Hermitian Hamiltonian \cite{Hsu2016, Chilcott2024},
\begin{equation}
	\bm{H}=
	\begin{pmatrix}
		E_1 & \Omega \\
		\Omega & E_2
	\end{pmatrix}
	- \frac{i}{2}
	\begin{pmatrix}
		\gamma_1 + \gamma'_1 & \sqrt{\gamma_1 \gamma_2}\\
		\sqrt{\gamma_1 \gamma_2} & \gamma_2 + \gamma'_2
	\end{pmatrix}
	\label{eq:two_level_hamiltonian}
\end{equation}
where the real term describes two bare levels with energies $E_1$ and $E_2$ coupled coherently by some external control with strength $\Omega$.
The imaginary term represents coupling to the continuum: the diagonal elements account for the individual decay of each level, while the off-diagonal elements $\sqrt{\gamma_1\gamma_2}$ describe coherent coupling between the two levels mediated by the shared continuum.
Coefficients $\gamma_{1,2}$ quantify the coherent part of the coupling, whereas $\gamma'_{1,2}$ represent additional incoherent decay paths that give rise to finite resonance widths.
This distinction is important because, unlike in a conventional Markovian bath, here the information transferred into the continuum can coherently return to the states, which is essential for the interferometric formation of a BIC.

The eigenvalues of Eq.~\ref{eq:two_level_hamiltonian} are generally complex, with their real and imaginary parts corresponding to the resonance positions and widths, respectively. 
As the detuning $\delta= E_1 - E_2$ changes sign, an avoided crossing forms.
Along this avoided crossing, the total resonance width, given by the sum of the imaginary parts of the eigenvalues, remains constant.
A special point is reached when the BIC condition
\begin{equation}
\delta_\text{BIC} = E_1 - E_2 = \Omega \frac{\gamma_1 - \gamma_2 + \gamma'_1 - \gamma'_2}{\sqrt{\gamma_1 \gamma_2}}
\label{eq:bic_condition}
\end{equation}
is fulfilled, at which one eigenstate acquires maximal and the other minimal width.
The state with minimal width corresponds to a coherent superposition of the two resonances, interfering such that the coupling to the continuum vanishes, closely analogous to the dark state in electromagnetically induced transparency (EIT), where a coherent superposition of ground states decouples from the excited state \cite{Fleischhauer2005}.

\section{Experimental Signature of the BIC}

Experimentally, we identify the BIC through its impact on the continuum wavefunctions.
The formation of a BIC is signaled by the simultaneous suppression of elastic and inelastic scattering interactions reflecting decoupling from the scattering continuum.
We probe inelastic two-body scattering via atom-loss spectroscopy, while elastic interactions are accessed through the analysis of collective excitations of the atomic cloud.
As an independent and complementary probe, we perform rf photoassociation, which directly probes the molecular wavefunction.

\subsection{Loss spectroscopy}

To study the avoided crossing and formation of the BIC, we prepare an ultracold sample of \Li at a temperature $T<\qty{600}{nK}$ inside a glass cell using techniques of optical and evaporative cooling (see Methods).
The atoms are prepared in an equal mixture of the two lowest hyperfine states, $\ket{a}=\ket{f=1/2, m_f = +1/2}$ and $\ket{b}=\ket{f=1/2, m_f = -1/2}$, where $f$ denotes the total angular momentum and $m_f$ its projection.
A static magnetic field $B_0$ and a time-periodic Floquet drive are then applied, and the remaining atom number is recorded after a $\qty{100}{ms}$ modulation pulse.
The Floquet modulation is generated by a pair of resonantly driven LC-circuits and characterized by frequency $\nu$ and amplitude $B_\text{rf} \approx \qty{8.5}{G}$.
For each spectroscopy data point, at least five measurements were averaged to reduce shot-to-shot fluctuations.
By scanning both $B_0$ and $\nu$, we observe a pronounced avoided crossing between two Floquet–Feshbach resonances $(I=0,\Delta_N=+1)$ and $(I=2,\Delta_N=0)$ [Fig.~\ref{fig:fig2}(a-c)], where $I$ labels the total nuclear spin of the underlying molecular state \cite{Chin2010}, and $\Delta_N$ the relative number of Floquet drive quanta \cite{Guthmann2025}.

Figure~\ref{fig:fig2}(d) shows that the coupling strength $\Omega$ between the two Floquet–Feshbach resonances increases linearly with modulation amplitude $B_\text{rf}$, demonstrating that Floquet modulation provides a precise handle to control the coupling between distinct Feshbach resonances.
At large modulation amplitudes, a quadratic AC Zeeman shift additionally shifts the center of the avoided crossing.
Notably, two frequency intervals can be identified where atom loss is strongly suppressed: region \rom{1} between $\qty{55.7}{}$ and $\qty{55.9}{MHz}$, and region \rom{2} between $\qty{58.1}{}$ and $\qty{58.9}{MHz}$.

To experimentally distinguish between regions \rom{1} and \rom{2}, we perform a sudden trap quench (details in Methods) and monitor the subsequent dynamical response of the atomic cloud [Fig.~\ref{fig:fig3}(a,b)].
This response provides a sensitive probe of elastic interactions \cite{PhysRevLett.92.203201, PhysRevLett.98.040401, Guthmann2025}.
In region \rom{2}, the cloud exhibits rapid thermalization and pronounced collective oscillations, indicating strong elastic interactions.
In contrast, in region \rom{1} no oscillations are observed.

To interpret these observations, we analyze the complex $s$-wave scattering length $a_s$, which characterizes low-energy collisions: its real part describes elastic interactions, whereas its imaginary part encodes inelastic loss \cite{Chin2010, hutson_feshbach_2007}.
Figure~\ref{fig:fig3}(c) shows $a_s$ and the corresponding loss spectra for several modulation frequencies across the avoided crossing.
In both regions \rom{1} and \rom{2}, the resonant feature of the imaginary part of $a_s$ becomes very small, explaining the strongly suppressed inelastic losses.
However, the real part behaves differently.
In region \rom{1} it becomes nearly featureless, demonstrating the decoupling of the resonance from the continuum and confirming that a BIC is formed.
By contrast, in region \rom{2} the real part remains strongly resonant, leading to enhanced elastic scattering despite the absence of inelastic loss.
A suppression of inelastic losses, analogous to what we observe here in region \rom{2}, was recently accomplished using a two-color excitation, but without an avoided crossing \cite{Guthmann2025}.

\begin{figure}
	\centering
	\includegraphics[scale=1.0]{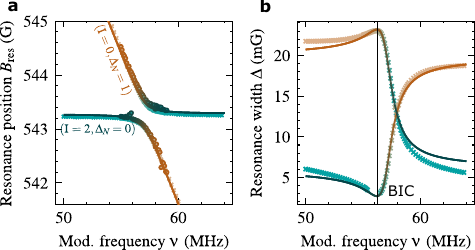}
	\caption{\textbf{Comparison with the non-Hermitian two-level model.} \textbf{(a)} Resonance positions extracted from the real parts of the eigenvalues of the effective non-Hermitian two-level Hamiltonian (solid line), compared with full coupled-channel calculations (crosses) and experimental data (dots). \textbf{(b)} Corresponding resonance widths derived from the imaginary parts of the eigenvalues (solid line) and from coupled-channel calculations (crosses). The BIC appears at the minimum of the resonance width. The small discrepancy between the predicted and experimentally observed BIC frequencies likely originates from uncertainties in the calibrated modulation amplitude and the underlying interaction potentials. The parameters of Eq.~\ref{eq:two_level_hamiltonian} were iteratively optimized to reproduce the coupled-channel results; details are provided in the Methods section. Colors indicate the admixture of the uncoupled bare states to the resulting eigenstates.}
    \label{fig:fig4}
\end{figure}

Figure~\ref{fig:fig4} provides a quantitative characterization of the Friedrich–Wintgen mechanism.
Panel (a) compares the resonance positions obtained experimentally and from full coupled-channel calculations with the real parts of the eigenvalues of the effective two-level Hamiltonian, Eq.~\eqref{eq:two_level_hamiltonian}.
Panel (b) shows the corresponding evolution of the resonance widths along the avoided crossing.
The BIC manifests as a pronounced minimum in the width of only $\qty{2.7}{mG}$, in stark contrast to the $\qty{17.0}{mG}$ width of the undressed $I=2$ Feshbach resonance from which it originates.

Because the sum of widths is conserved along the avoided crossing \cite{Friedrich1985}, the complementary resonance correspondingly acquires maximal width, sometimes referred to as superradiant state \cite{Sadreev2021}.
Experimentally, the measured atom losses reflect inelastic processes; thus, the elastic width is extracted from coupled-channel calculations.
A small residual width persists at the BIC point due to coupling to higher-order Floquet continua, an effect incorporated phenomenologically through the incoherent decay coefficients $\gamma'_{1,2}$ in Eq.~\ref{eq:two_level_hamiltonian}.
Details of the optimization procedure used to match the effective model to our system are provided in the Methods section.

\subsection{RF Photoassociation}
To further investigate how the interference of Feshbach resonances modifies the scattering wavefunction and causes the formation of a BIC, we employ the tool of rf photoassociation.
Photoassociation from a scattering continuum of free atoms serves as a sensitive probe of molecular wavefunctions, since the line strength is governed by the Franck–Condon wave-function overlap between the free-atom scattering state and the molecular state \cite{Jones2006, STWALLEY1999194, Ding2017}.

For this purpose, we prepare a cold atomic sample in a mixture of hyperfine states $\ket{a}$ and $\ket{c}=\ket{f=3/2, m_f=-1/2}$. 
In addition to the Floquet modulation, we apply a second rf pulse with frequency $\nu_\text{spec}$ and polarization perpendicular to the quantization axis, which drives spin-flip transitions from $\ket{c}$ to $\ket{b}$.

\begin{figure}
	\centering
	\includegraphics[scale=1.0]{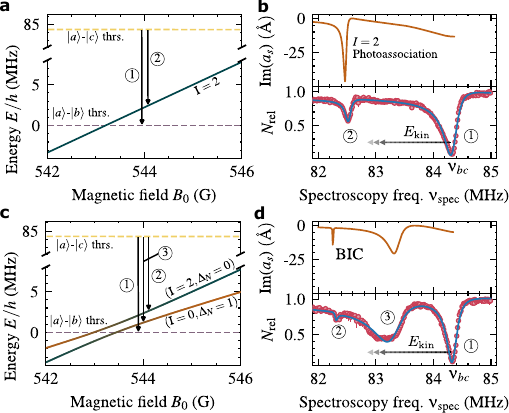}
	\caption{\textbf{RF photoassociation spectroscopy.} \textbf{(a)} Energy-level diagram and working principle of RF photoassociation. Dashed lines denote atomic thresholds (thrs.), while solid lines indicate molecular bound states below threshold and scattering states above. The gas is prepared in a $\ket{a}$-$\ket{c}$ mixture, and the RF field drives the $\ket{c}\rightarrow\ket{b}$ transition (arrow (1)). Besides single-atom spin flips, two colliding atoms in the $\ket{a}$-$\ket{c}$ channel can be photoassociated into $\ket{a}$-$\ket{b}$ molecules (arrow (2)), with a rate set by the Franck–Condon overlap of the scattering states. \textbf{(b)} Atom-loss spectrum by rf photoassociation (lower panel) and imaginary part of $a_s$ (upper panel) without Floquet modulation. The RF detuning $\nu_\delta$ defines the probed kinetic energy $E_\text{kin}$. Two loss features appear: (1) the single-atom spin-flip transition and (2) a two-body rf photoassociation resonance linked to the static $I=2$ Feshbach resonance. \textbf{(c)} With Floquet modulation of $\qty{55.74}{MHz}$, the dressed $(I=0,\Delta_N=+1)$ resonance intersects the static $(I=2,\Delta_N=0)$ resonance at the $\ket{a}$-$\ket{b}$ threshold. \textbf{(d)} Corresponding spectrum showing the atomic spin-flip resonance (1) and two photoassociation features: the original narrow (2) and a broad (3) resonance. (3) acquires the full width, while (2) becomes narrow, marking the formation of a BIC. The deviation of the fit from the data for (3) might be caused by a mixture of energy broadening and a varying rabi frequency with spectroscopy frequency. A slight position shift can be observed due to the coupling between the two resonances. }
	\label{fig:fig5}
\end{figure}

By tuning $\nu_\text{spec}$ below the $\ket{c}\rightarrow\ket{b}$ transition frequency $\nu_{bc}$, we access the scattering wavefunction in the $\ket{a}$-$\ket{b}$ channel.
The detuning $\nu_\delta = \nu_{bc} - \nu_\text{spec}$ directly corresponds to the kinetic energy $E_\text{kin} = h \nu_\delta$ above the $\ket{a}$-$\ket{b}$ threshold at which the wavefunction is probed, as illustrated in Fig.~\ref{fig:fig5}.

In Fig.~\ref{fig:fig5}(a,b), the energy levels and the corresponding photoassociation spectra in the absence of Floquet modulation at a static magnetic field of $B_0 = \qty{544.0}{G}$ are shown. 
The photoassociation spectrum exhibits two features: a dominant dip at $\qty{84.35}{MHz}$ corresponding to the single-atom spin-flip transition $\nu_{bc}$, and a weaker dip at $\qty{82.6}{MHz}$ caused by rf photoassociation to the $I=2$ Feshbach resonance state.

With Floquet modulation applied [Fig.~\ref{fig:fig5}(c,d)], the energy levels and photoassociation spectra are modified. 
For a Floquet modulation frequency $\nu=\qty{55.74}{MHz}$ and drive strength $B_\text{rf}=\qty{8}{G}$, the dressed $(I=0, \Delta_N=+1)$ resonance forms an avoided crossing with the $(I=2, \Delta_N=0)$ resonance above the $\ket{a}$-$\ket{b}$ threshold.
Coupled-channel calculations under these conditions predict the emergence of a BIC at a kinetic energy of approximately $\qty{2}{MHz}\times h \approx \qty{100}{\mu K}$ above threshold.

The resulting photoassociation spectrum exhibits three distinct features: the single-atom spin-flip transition and two rf photoassociation resonances.
With the Floquet drive enabled, one of these association resonances is strongly suppressed compared to the undriven case [Fig.~\ref{fig:fig5}(b)], while the other correspondingly acquires the full resonance width.

This suppression reflects the strongly reduced wavefunction amplitude of the BIC in the open $\ket{a}$-$\ket{b}$ channel, confirming decoupling from the scattering continuum.

\section{Interference protected Quantum States in an Open System}

The ability to engineer a BIC through Floquet control provides a powerful means to manipulate both interactions and dissipation in ultracold matter.
Such interference-protected states could serve as long-lived gateways for ground-state molecule formation via Raman or STIRAP techniques \cite{Vitanov2017}, or as a controllable platform for exploring non-Hermitian and topological quantum dynamics, including exceptional points \cite{Bergholtz2021, Xu2016}.

An immediate extension of this work is to investigate whether the residual width of the BIC can be further reduced using multi-color Floquet drives, a method that has already proven effective in suppressing inelastic losses of Floquet–Feshbach resonances \cite{Guthmann2025} and driven Floquet–Hubbard systems \cite{Viebahn2021}.
Beyond this, the tunable interplay between coherent coupling and controlled dissipation demonstrated here establishes a foundation for programmable open quantum systems and the exploration of interference-stabilized driven quantum matter.

\begin{acknowledgments}
    We thank Akash Kamra, André Eckardt, Julian Fess, Yifei Bai, Andreas Guthmann and Aurélien Perrin for providing insightful comments and carefully reading our manuscript. This work was supported by the German Research Foundation (DFG) via the Collaborative Research Center Sonderforschungsbereich SFB/TR185 (project 277625399) and the European Research Council (ERC) under the European Union’s Horizon Europe research and innovation programme (ERC Advanced Grant No. 101200776).
\end{acknowledgments}

\bibliography{bibliography}

\section{Methods}

\subsection{Experimental Details}
\begin{figure}
	\centering
	\includegraphics[scale=0.95]{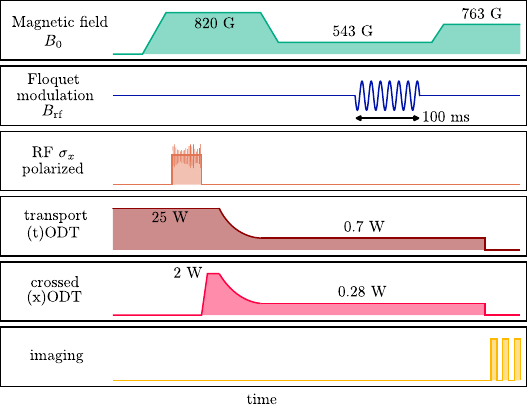}
	\caption{\textbf{Experimental loss spectroscopy sequence.} After transport into the glass cell, the magnetic field is rapidly ramped to \qty{820}{G}, where a spectrally broadened RF pulse is applied to prepare an equal mixture of $\ket{a}$ and $\ket{b}$. The gas is then evaporatively cooled in a crossed dipole trap to temperatures below \qty{600}{\nano\kelvin}, after which the magnetic field is ramped to the target value. Once the field has stabilized, the Floquet modulation with variable frequency is applied for \qty{100}{\milli\second}. Finally, the magnetic field is ramped back to \qty{763}{G}, and the remaining atoms are detected by resonant absorption imaging.}
	\label{fig:fig_expseq}
\end{figure}

We begin by preparing a two-component thermal gas of $^6$Li atoms in the two lowest hyperfine states, $\ket{a}$ and $\ket{b}$, confined in an optical dipole trap inside a steel vacuum chamber \cite{Gaenger2018}.
The trapped gas is then transported into a glass cell using a focus-tunable lens.
Performing the experiment inside the glass cell is essential to achieve the required modulation depth of the oscillating magnetic field.
A schematic overview of the experimental sequence is shown in Fig.~\ref{fig:fig_expseq}.
After transfer, a short spectrally broadened RF pulse is applied to ensure an equal spin mixture of $\ket{a}$ and $\ket{b}$.
The magnetic field is then ramped to \qty{820}{G}, where a second dipole trap is switched on forming a crossed dipole trap, enabling efficient evaporative cooling.
Evaporation is terminated just before molecule formation sets in, resulting in a temperatures below \qty{600}{nK}.
The ultracold cloud is subsequently brought to the target magnetic field near \qty{543}{G}, where it is held for \qty{200}{ms} to allow the field to stabilize, resulting in the faint horizontal feature of the static $I=2$ Feshbach resonance visible in \autoref{fig:fig2}(a).
The Floquet modulation is then applied for \qty{100}{ms}, generated by two resonantly driven LC circuits \cite{Guthmann2025}.
After modulation, the magnetic field is ramped back to \qty{760}{G}, and the remaining atoms are detected via resonant absorption imaging on the D$_2$ transition.

To probe the elastic interaction strength, we perform trap-quench measurements as described in Ref. \cite{Guthmann2025}.
The gas is prepared following the procedure above, except that evaporation is performed only in the transport trap, without the crossed-beam configuration.
During the application of the Floquet drive, the trap power is suddenly reduced from \qty{1}{W} to \qty{200}{mW}, exciting collective breathing dynamics.
After a variable evolution time, the Floquet drive is switched off and the cloud is imaged immediately.

\begin{figure}
	\centering
	\includegraphics[scale=0.95]{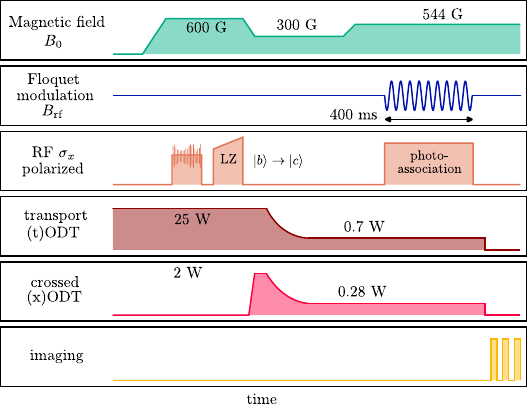}
	\caption{\textbf{Experimental rf photoassociation sequence.} After preparing the ultracold gas in the glass cell (see Fig.~\ref{fig:fig_expseq}), the magnetic field is ramped to \qty{600}{G}, where a spectrally broadened rf pulse is applied. A subsequent Landau–Zener sweep transfers atoms from $\ket{b}$ to $\ket{c}$, producing an $\ket{a}$-$\ket{c}$ mixture. This mixture is further evaporatively cooled at \qty{300}{G} before the field is ramped to the target value of \qty{544}{G}. At this field, a high-power rf pulse drives photoassociation from the $\ket{a}$-$\ket{c}$ continuum to the $\ket{a}$-$\ket{b}$ channel, while the Floquet modulation is applied simultaneously to access the BIC. Detection is performed using the same absorption imaging procedure as in the loss-spectroscopy sequence.}
	\label{fig:fig_expseqrf}
\end{figure}

For the rf photoassociation measurements, the sequence is slightly modified (see Fig.~\ref{fig:fig_expseqrf}).
After preparing an equal mixture of $\ket{a}$ and $\ket{b}$, the magnetic field is ramped to \qty{600}{G}, where a Landau–Zener sweep transfers atoms from $\ket{b}$ to $\ket{c}$.
The resulting $\ket{a}$-$\ket{c}$ mixture is then evaporatively cooled at \qty{300}{G}, before the magnetic field is ramped to the target value of \qty{544}{G}.
After allowing the magnetic field to stabilize, a high-power $\sigma_x$-polarized RF pulse with a Rabi frequency of $2\pi\times\qty{81}{\kilo\hertz}$ is applied for a duration of \qty{400}{\milli\second} to drive the photoassociation transition from the continuum of $\ket{a}$-$\ket{c}$ scattering states to the $\ket{a}$-$\ket{b}$ continuum.
To observe the BIC, Floquet modulation polarized along the quantization axis is applied simultaneously.
The remaining atoms are detected using the same resonant absorption imaging method described above.

The Floquet modulation depth $B_\text{rf}$ is calibrated via the AC-Zeeman shift of the hyperfine levels induced by the drive, following the procedure in Ref.~\cite{Guthmann2025}.

\subsection{Coupled-Channel Calculations}
At ultracold temperatures, two-body collisions are dominated by $s$-wave scattering and can be characterized by the complex $s$-wave scattering length $a_s$ .
For purely elastic collisions, $a_s$ is real, while inelastic processes introduce an imaginary component that accounts for loss.
The elastic and inelastic scattering cross sections follow directly from $a_s$ \cite{cvitas2007}.

To compute $a_s$ we employ coupled-channel methods, which are well suited for cold-atom systems where multiple internal channels contribute to the scattering process.
These methods solve the full multichannel Schrödinger equation, either in its differential form or equivalently as a set of coupled Lippmann–Schwinger integral equations.
We solve the coupled Lippmann–Schwinger equations with an algorithm that expands the scattering wavefunction in Chebyshev polynomials to obtain accurate and numerically stable results \cite{Rawitscher_2005, GONZALES1997134, GONZALES1999160}.

Accurate coupled-channel calculations require precise molecular interaction potentials.
The potentials used in this work are based on high-resolution spectroscopic data \cite{le_roy_accurate_good, DATTANI2011199} and were further refined following the procedure of Ref. \cite{julienne_contrasting_good}.
Details of the Floquet Hamiltonian used for the numerical results presented here are given in Ref. \cite{Guthmann2025}.
For the calculation of $\text{Im}(a_s)$ relevant to the RF photoassociation data shown in \autoref{fig:fig5}(d), we extend the Hamiltonian to include an additional field term that accounts for the $\sigma_x$-polarized RF spectroscopy drive.

For the thermally averaged inelastic cross section shown in \autoref{fig:fig2}(c), we compute $\sigma_\text{inel}(E_c)$ for several collision energies $E_c$ and perform a Boltzmann-weighted average.
With $p(E_c)$ the Boltzmann distribution, the averaged cross section is

\begin{equation}
\sigma_\text{inel}^\text{avg}
= \int p(E_c) \sigma_\text{inel}(E_c) \mathrm{d}E_c \: .
\end{equation}

\subsection{Parameters of the two-level Hamiltonian}

To adapt \autoref{eq:two_level_hamiltonian} for modeling the avoided crossing between the $(I=2,\Delta_N=0)$ and $(I=0,\Delta_N=+1)$ Floquet–Feshbach resonances, the bare levels $E_1$ and $E_2$ must be redefined.
We assign $E_2$ to the static $(I=2,\Delta_N=0)$ resonance, which is not Floquet-dressed and therefore independent of the modulation frequency.
Its energy is given by $E_2= \mu_B B_0$, where $B_0$ denotes the magnetic-field position of the conventional static $I=2$ Feshbach resonance and $\mu_B$ the Bohr magneton.
In contrast, the $(I=0,\Delta_N=+1)$ resonance assigned to $E_1$ depends on the driving frequency $\nu$. 
Although this dependence is generally nonlinear \cite{Guthmann2025}, it can be well approximated as linear within the experimentally relevant frequency range.
Accordingly, we model it as $E_1 = \mu_B [B_0 + \kappa(\nu - \nu_0)]$, where $\nu_0$ is the modulation frequency at which the two Floquet-Feshbach resonances would intersect at $B_0$ in the absence of coupling $\Omega$.
With these definitions, the Hamiltonian \autoref{eq:two_level_hamiltonian} becomes
\begin{equation}
    \bm{H}=
	\begin{pmatrix}
		\mu_B [B_0 + \kappa(\nu - \nu_0)] & \Omega \\
		\Omega & \mu_B B_0
	\end{pmatrix}
	- \frac{i}{2}
	\begin{pmatrix}
		\gamma_1 + \gamma'_1 & \sqrt{\gamma_1 \gamma_2}\\
		\sqrt{\gamma_1 \gamma_2} & \gamma_2 + \gamma'_2
	\end{pmatrix}
    \label{eq:magnetic_hamiltonian}
\end{equation}
from whose complex eigenvalues $\lambda_{1/2}$ the resonance positions and widths are obtained via division by $\mu_B$.

The parameters of \autoref{eq:magnetic_hamiltonian} were determined through an iterative fitting procedure.
First, the real part of the Hamiltonian was optimized so that $\text{Re}(\lambda_{1/2})$ reproduced the experimentally observed resonance positions.
Next, the imaginary part was adjusted to match $\text{Im}(\lambda_{1/2})$ to the numerically calculated resonance widths.
Because modifying the imaginary part slightly shifts the resonance positions, both steps were alternated until convergence was achieved.
The resulting optimized parameters are listed in \autoref{tab:parameters}.

\begin{table}[h]
\centering
\caption{Parameters of \autoref{eq:magnetic_hamiltonian} optimized to fit the observed avoided crossing between the $(I=2,\Delta_N=0)$ and $(I=0,\Delta_N=+1)$ Floquet–Feshbach resonances.}
\begin{tabular}{ |c|c| }
 \hline
Parameter & Value  \\ 
 \hline
 $B_0$ & \qty{543.28}{G}  \\ 
 $\kappa$ & \qty{-0.5443}{G / \mega\hertz}  \\ 
 $\nu_0$ & \qty{57.0}{\mega\hertz}  \\ 
 $\Omega / h$  & \qty{327.79}{\kilo \hertz}  \\ 
 $\gamma_1 / h$ & \qty{48.69}{\kilo \hertz}  \\ 
 $\gamma_2 / h$ & \qty{9.18}{\kilo \hertz}  \\ 
 $\gamma'_1 / h$ & \qty{6.93}{\kilo \hertz}  \\ 
 $\gamma'_2 / h$ & \qty{7.70}{\kilo \hertz}  \\ 
 \hline
\end{tabular}
\label{tab:parameters}
\end{table}

\subsection{Fitting of loss spectroscopic data}
To fit the experimental loss data we use a sum of Fano profiles \cite{PhysRev.137.A1364},
\begin{equation}
    N_\text{rel} B = \sum_i N_i \frac{(q_i + \beta_i)^2}{1 + \beta_i^2} + N_0 \: ,
\end{equation}
with the shape parameter $q$ and
\begin{equation}
    \beta_i = \frac{B - B^{(0)}_i}{\Delta_i /2} \: ,
\end{equation}
where $B^{(0)}_i$ corresponds to the position and $\Delta_i$ to width of the $i$-th resonance.
The factor $N_0$ accounts for a constant offset and $N_i$ scales the height of individual resonances.

From the loss spectroscopic data the magnetic splitting $\Delta_B=(B^{(0)}_1-B^{(0)}_2)$ was extracted and from this the coupling strength via $2\Omega=\mu_B \Delta_B$ to obtain \autoref{fig:fig2}(d).

\subsection{Fitting of quench response}
To quantify the elastic scattering contribution shown in Fig.~\ref{fig:fig3}(a,b), we analyze the cloud’s response to a sudden trap quench following the procedure of Ref.~\cite{Guthmann2025}.
At each evolution time, several absorption images were recorded and averaged, and the transverse density profile was fitted with a Gaussian to extract the cloud width $\sigma(t)$.
The resulting time series was then modeled by
\begin{equation}
\sigma(t)= \sigma_0
+ A e^{-\gamma t}\cos\left(\omega t + \theta\right)
+ B e^{-\gamma_B t},
\end{equation}
where $\sigma_0$ denotes the equilibrium width in the absence of oscillations.
The oscillatory term describes the breathing-mode response of the trapped gas with amplitude $A$, damping rate $\gamma$, frequency $\omega$ and phase $\theta$.
The final exponential captures non-oscillatory relaxation processes with amplitude $B$ and decay rate $\gamma_B$.
Plotting $A/\gamma$ reveals the elastic resonance response.

\end{document}